\documentclass[runningheads]{llncs}
\usepackage[hidelinks]{hyperref}

\usepackage{epsfig,endnotes}
\usepackage{algorithm}
\usepackage{algpseudocode}
\usepackage{subcaption}
\usepackage{amsmath}
\usepackage{stfloats}
\usepackage[export]{adjustbox}
\usepackage{xcolor}
\usepackage{tabularx}
\usepackage{ragged2e}
\usepackage{multirow}
\usepackage{makecell}
\usepackage{booktabs} 
\usepackage[justification=centering]{caption}
\usepackage[usenames,dvipsnames]{xcolor}
\usepackage{url}
\usepackage{placeins}
\usepackage{graphicx}
\usepackage[hidelinks]{hyperref}

\newcommand{\orcidicon}[1]{%
  \unskip\kern0.1em%
  \href{https://orcid.org/#1}{%
    \includegraphics[width=8pt]{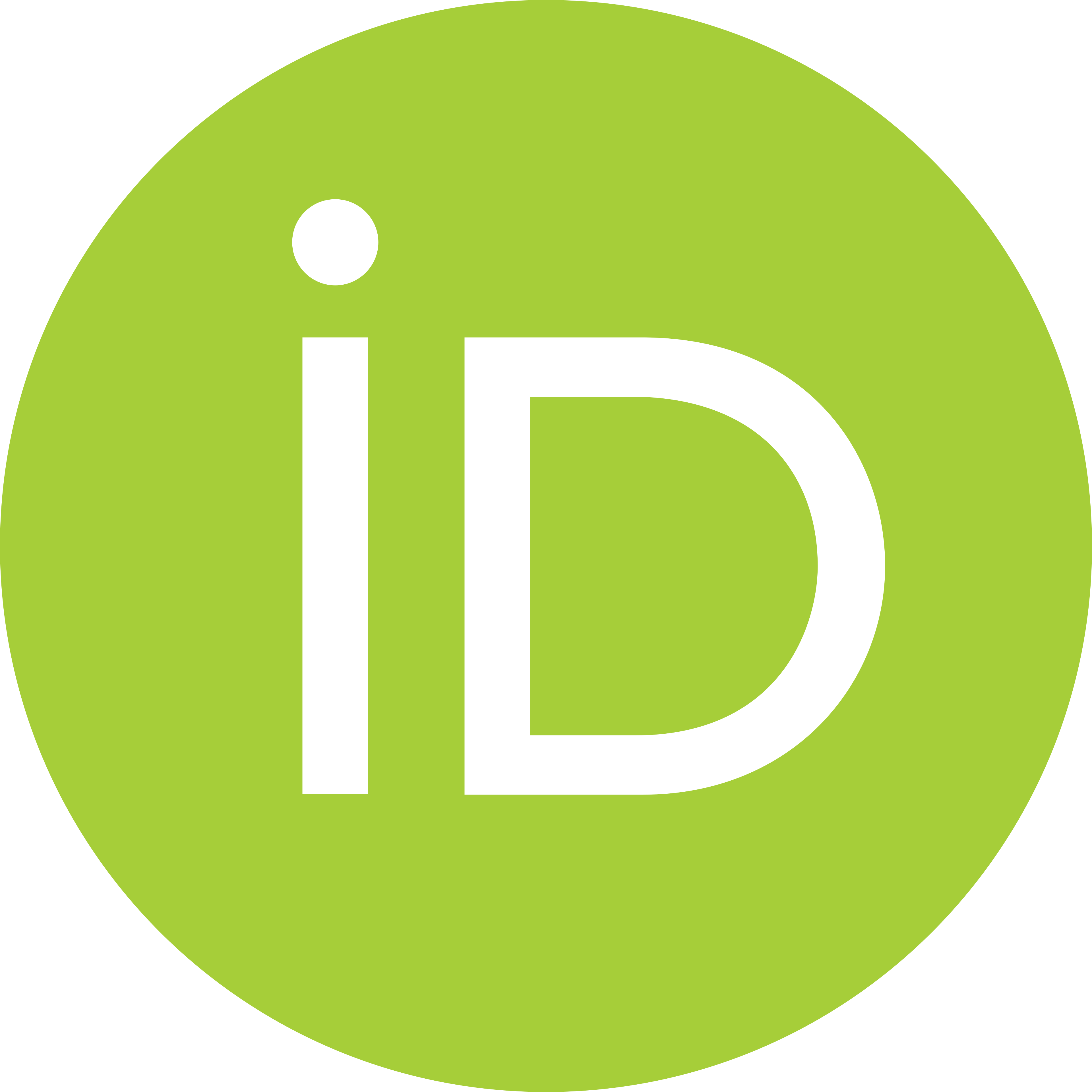}% 
  }%
}

\begin{document}
\title{Dodoor: Efficient Randomized Decentralized Scheduling with Load Caching for Heterogeneous Tasks and Clusters}
\titlerunning{Dodoor}
%
%\titlerunning{Abbreviated paper title}
% If the paper title is too long for the running head, you can set
% an abbreviated paper title here
%
\author{Wei Da\inst{1}\orcidicon{0009-0003-7305-6531}\thanks{Corresponding author.} \and
Evangelia Kalyvianaki\inst{1}\orcidicon{0000-0003-0753-1261}}
\authorrunning{W. Da and E. Kalyvianaki}
%
% First names are abbreviated in the running head.
% If there are more than two authors, 'et al.' is used.
%
\institute{University of Cambridge, Cambridge, United Kingdom \\
\email{wd312@cam.ac.uk}, \email{ek264@cam.ac.uk}}
\maketitle              % typeset the header of the contribution
\begin{abstract}
This paper presents Dodoor, a randomized decentralized scheduler for heterogeneous clusters. Dodoor removes hot-path probing via batched cache refreshes and introduces a heterogeneity-aware resource-load score that ranks sampled candidates using multidimensional fit and queued-duration pressure. On a 101-node CloudLab cluster, Dodoor cuts scheduler messages by 55--66\% while improving throughput by up to 33.2\% (Azure) and 21.5\% (FunctionBench) compared with decentralized baseline schedulers. Dodoor also reduces mean makespan latency by up to 12.1\% and 7.2\%, and tail latency by up to 21.9\% and 24.6\%.

\end{abstract}
\section{Introduction}
Modern datacenters require schedulers that are both fast and resource-aware under sustained high request rates. In practice, this is challenging because workloads and clusters are heterogeneous as task demands vary across resource types, and servers expose different capacity and performance profiles. In this setting, placement quality and scheduling overhead are tightly coupled. Centralized schedulers can optimize with cluster-wide states, but maintaining and synchronizing this state at high request rates introduces overheads~\cite{omega,apollo,quincy,borg}. Decentralized randomized schedulers based on the power-of-two model scale naturally by allowing replicas to schedule independently across sampled candidate nodes~\cite{sparrow,envoy,f5NGINXPower,prequal}. However, many decentralized implementations still rely on per-request probing, making control-plane RPC traffic a first-order bottleneck.

This overhead–quality tradeoff further grows in heterogeneous clusters. Many practical decentralized schedulers rank candidates by their local queue length, which is often a weak proxy for placement quality. A node with a short queue may still be a poor fit for a task's multidimensional demand, while a slightly busier node may provide better residual-capacity fit~\cite{tetris,head-of-line-tcp}.

% With concurrent schedulers and stale views on the nodes, probe-based placement can still herd schedulers onto the same seemingly lightly loaded node, increasing contention and tail latency.

We present Dodoor, a decentralized scheduler for heterogeneous clusters that makes probe-free placements from cluster cached state while preserving heterogeneity-aware decisions. Dodoor separates the control plane into schedulers, workers, and a lightweight data service; schedulers place tasks from local cached state; workers and schedulers emit compact deltas; and the data service pushes batched snapshots back to schedulers. Also, Dodoor retains low-overhead pairwise candidate sampling but replaces queue-only ranking with a resource-load score that combines multidimensional fit and queued-duration pressure.

Grounded in batched balls-into-bins theory~\cite{batchSetting,power_of_two_batches}, Dodoor targets decentralized early-binding in heterogeneous clusters, where full global synchronization is undesirable but naive queue-based randomness is insufficient. We evaluate Dodoor on a 101-node heterogeneous CloudLab deployment with two workloads: replayed Azure VM traces and containerized FunctionBench Python tasks. Across both, Dodoor reduces scheduler communication by 55--66\% and improves end-to-end performance over representative decentralized baselines, with throughput gains up to 33.2\% (Azure) and 21.5\% (FunctionBench) alongside latency reductions.

The contributions of this paper are: (1) a probe-free decentralized scheduling path that combines batched push-based cache refresh with randomized pairwise candidate selection, reducing communication and scheduler herding under stale local views; (2) a heterogeneity-aware resource-load score that jointly models multidimensional task demand, residual server capacity, and queued-duration pressure; and (3) a real-cluster evaluation (101 nodes; Azure VM trace + FunctionBench containers) that quantifies communication, throughput, and latency benefits and tradeoffs.

\section{Background and Motivation} \label{motivation}
This section summarizes the theoretical and systems background that motivates Dodoor. We first review the balls-into-bins problem, then discuss how randomized schedulers are built on these ideas, and finally we clarify Dodoor's motivation, scope, and contribution.

\subsection{Balls-into-bins Problem} \label{theory}
The balls-into-bins problem models assigning $m$ tasks (balls) to $n$ servers (bins), typically with $m\gg n$. With a simple random scheduler, each task is assigned to one uniformly random server, which gives a load gap of $\Theta(\sqrt{(m\log n)/n})$ above the average~\cite{power_of_two_raw}. The classic power-of-$d$ process samples $d$ servers and picks the least loaded one; for $d=2$ this dramatically improves concentration and yields the power-of-two behavior used in real systems.

Beyond the uniform setting, several variants have been studied. In the \emph{weighted} model, balls have non-uniform weights and bin load is defined as the sum of assigned weights~\cite{balancedWeightedCase,batchSetting}. Load information can also be partial, stale, or delayed~\cite{stale_load,how_useful_of_old,batchSetting}. A key special case is the $b$-\emph{batched} setting, where load updates arrive once per batch of size $b$. Combining batching with weights, prior work further shows that weighted batched power-of-two still admits provable and controllable imbalance bounds~\cite{power_of_two_batches}.

\subsection{Power-of-two in Modern Schedulers} \label{section:randomized_scheduler}
Datacenter schedulers are often described as centralized versus decentralized. Centralized schedulers could make optimal decisions based on a rich global view, but they pay higher synchronization and metadata costs as clusters and request rates grow~\cite{quincy,Tetrisched,omega,apollo}. Decentralized power-of-two schedulers scale better because each scheduler can act independently on a small sampled state.

This pattern is widely adopted in practical systems and proxies (e.g., Nginx- or Envoy-like designs) that probe a small candidate set of nodes and dispatch to the least-loaded node~\cite{f5NGINXPower,envoy}. However, these implementations usually approximate load by queue length or requests in flight. Under heterogeneous CPU/memory capacities, ``least queue'' can differ substantially from ``best fit,'' causing fragmentation and head-of-line blocking effects~\cite{tetris,head-of-line-tcp}. Moreover, when concurrent schedulers see similarly stale signals, heavy probe overlap can trigger synchronized choices toward the same apparently light nodes. This thundering-herd effect amplifies tail latency and increases control-plane traffic at high QPS.

\begin{figure}[t]
    \centering
    \includegraphics[width=0.7\columnwidth]{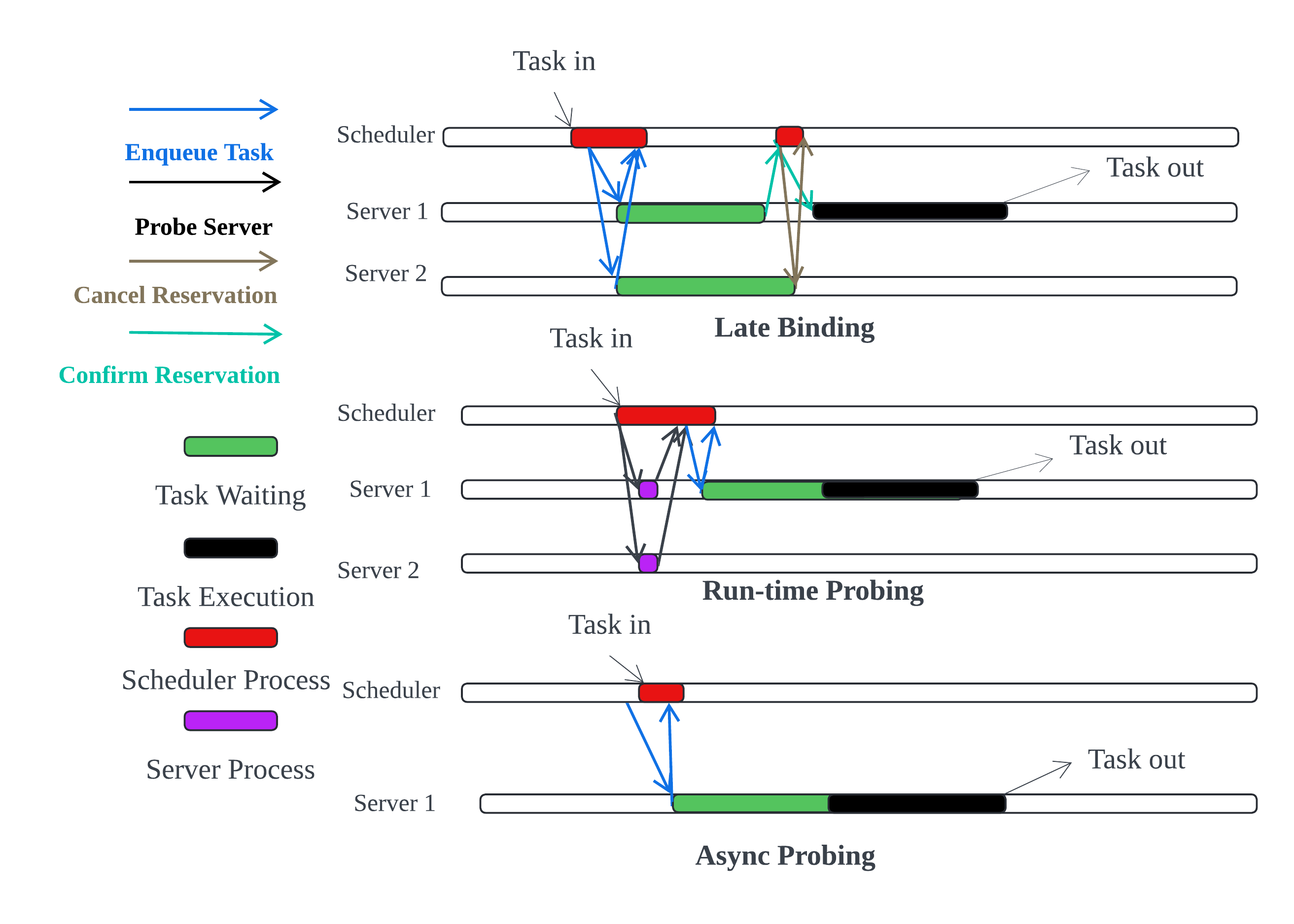}
    \caption{Probe-based scheduling message flow (single request path).}
    \label{fig:probing_strategy}
\end{figure}

\subsection{Variants of Randomized Schedulers} \label{section:variants_of_random}
Standard power-of-two with runtime probing remains the most common approach in production systems. In addition, Sparrow-style late binding reserves multiple candidates and later confirms or cancels placement based on queue progress~\cite{sparrow}. It is designed for slot-oriented, relatively homogeneous settings, but is less suited to practical settings where tasks and servers have varying resource demands and capacities. In particular, late binding introduces an admission-confirmation gap as a node that looks reservable may still be inadmissible at commit time under multidimensional resource constraints. Under concurrent schedulers, this gap increases confirmation races, retry/cancel traffic, and node reservation complexity.

Prequal keeps asynchronous sampled observations in a scheduler-local pool and reuses them across future requests~\cite{prequal}. By contrast, YARP primarily relies on periodic synchronized probing/health polling at fixed intervals to refresh backend state~\cite{microsoftYARPDocumentation}. Both approaches reduce per-request probe work but introduce a freshness-management tradeoff: stale or coarsely refreshed probe results can hurt decision quality, while aggressive refresh/removal raises control overhead. More broadly, these decentralized schedulers still mainly optimize scalar queue signals rather than multidimensional resource fit. Thus, existing randomized schedulers trade freshness against communication, motivating batched push-based state dissemination and a richer heterogeneity-aware load metric. Figure~\ref{fig:probing_strategy} summarizes these probing strategies' message flows.

\subsection{Scope and Proposal} \label{section:scope}
Dodoor provides decentralized, heterogeneity-aware placement for clusters, and eliminates per-request probing on the hot-path. Concretely, Dodoor maps weighted and batched insights from balanced allocation into a practical architecture where workers and schedulers send compact deltas to a data service, and schedulers receive batched snapshots for local decisions. Dodoor follows an asynchronous state-dissemination model, but unlike Prequal/YARP, it is probe-free on the scheduling path, which leads to overhead reduction and reliability improvement. At the scheduling-algorithm level, Dodoor keeps low-overhead power-of-two candidate sampling but replaces queue-only ranking with a task-aware resource-load score that combines multidimensional resource fit and queued-duration signals. This design directly addresses the gap identified above as it retains decentralized scalability while improving decision quality for heterogeneous tasks and servers.

Dodoor does not claim a new balls-into-bins theorem. Rather, its novelty is the architectural synthesis of batched push-based cache refresh, power-of-two randomization under stale views, and a heterogeneity-aware resource-load (RL) score. This combination removes per-request probing from the hot path, lets schedulers operate fully in parallel without runtime synchronization, and keeps strict global consistency off the critical path. To our knowledge, Dodoor is among the first decentralized early-binding schedulers to combine these elements in a single architecture with real-cluster validation.

\section{Dodoor Scheduling} \label{scheduler}
\subsection{System Model} \label{system_model}
We model a heterogeneous cluster with a set of workers $\mathcal{N}=\{1,\dots,n\}$ that run the tasks, and a set of schedulers $\mathcal{S}=\{1,\dots,m\}$ that make scheduling decisions. Each worker $i\in\mathcal{N}$ has a $K$-dimensional capacity vector $\vec{C}_i\in\mathbf{R}_{+}^{K}$ (e.g., CPU and memory). For each incoming task $t$, scheduler $s$ observes demand vector $\vec{r}_t\in\mathbf{R}_{+}^{K}$ and per-server duration estimate $\hat{d}_{t,i}$ (if duration is server-agnostic, $\hat{d}_{t,i}=d_t$ for all $i$). Besides, Dodoor includes a lightweight data service that asynchronously aggregates updates and pushes refreshed cluster snapshots.

Scheduler replica $s\in\mathcal{S}$ stores local cached state $\{\mathcal{T}_{i,s},\vec{L}_{i,s}, D_{i,s}\}_{i=1}^{n}$, where $\mathcal{T}_{i,s}$ is the cached queued/running task set for server $i$,
\[
\vec{L}_{i,s}=\sum_{u\in\mathcal{T}_{i,s}}\vec{r}_u,
\qquad
D_{i,s}=\sum_{u\in\mathcal{T}_{i,s}}\hat{d}_{u,i},
\]
i.e., $\vec{L}_{i,s}$ is the sum of resource-demand vectors of uncompleted tasks on worker $i$ and $D_{i,s}$ is the sum of their estimated durations on $i$. The data service pushes refreshed snapshots to schedulers every global batch of size $b$ (counted in scheduling decisions), and each scheduler flushes an aggregated delta every $b_{\Delta}$ placements; we typically set $b_{\Delta}\le 2b/|\mathcal{S}|$ so all schedulers contribute updates before the next push. Section~\ref{section:scheduler_service} details the API and message flow.

For each candidate server $i$, the RL (resource-load) affinity score is defined as
\begin{equation}
rl_{i,t,s}=\frac{\vec{r}_t^\top\vec{L}_{i,s}}{\lVert\vec{C}_i\rVert_2^2},
\label{eq:rl}
\end{equation}
where $\lVert\vec{C}_i\rVert_2^2=\sum_{\kappa=1}^{K} C_{i,\kappa}^2$ is the normalization of server capacity. For a sampled pair $a,c\in\mathcal{N}$, let $\tilde{D}_a=D_{a,s}+\hat{d}_{t,a}$ and $\tilde{D}_c=D_{c,s}+\hat{d}_{t,c}$. Dodoor computes
\begin{equation}
\begin{aligned}
score_a&=(1-\alpha)\frac{rl_a}{rl_a+rl_c}+\alpha\frac{\tilde{D}_a}{\tilde{D}_a+\tilde{D}_c},\\
score_c&=(1-\alpha)\frac{rl_c}{rl_a+rl_c}+\alpha\frac{\tilde{D}_c}{\tilde{D}_a+\tilde{D}_c},
\end{aligned}
\label{eq:pair-score}
\end{equation}
and selects the lower-score candidate.

\subsection{Scheduling Algorithm} \label{scheduling_algo}
Algorithm~\ref{alg:scheduling} summarizes the online scheduling algorithm executed by each scheduler. The scheduling process has two stages: randomized pair sampling and heterogeneity-aware pair scoring. The weight $\alpha\in[0,1]$ controls the tradeoff between resource balancing ($\alpha\to0$) and queue-duration sensitivity ($\alpha\to1$). Here $b$ controls data-service-to-scheduler cache pushes, while $b_{\Delta}$ controls scheduler-to-data-service delta flushes ($b_{\Delta}<b$ in practice). This separation keeps the scheduler's hot path probe-free and prevents worker-side state from staying stale for too long. The predicate $\textproc{newCacheAvailable}()$ returns true once the data service has delivered a snapshot newer than the scheduler's current cache, so cache refresh is event-driven rather than polled.

\begin{algorithm}[t]
    \caption{Dodoor Pairwise Placement (Scheduler $s$, Task $t$)}
    \label{alg:scheduling}
    \begin{algorithmic}[1]
    \Require cache $(\vec{L}_{\cdot,s}, D_{\cdot,s})$, capacities $\vec{C}$, cache batch $b$, delta batch $b_{\Delta}$, weight $\alpha$
    \State sample two distinct candidates $a,c\sim\textproc{Uniform}(\mathcal{N})$
    \State compute $(score_a,score_c)$ by Eqs.~\eqref{eq:rl}--\eqref{eq:pair-score}
    \State $j\gets a$ if $score_a\le score_c$, else $j\gets c$
    \State enqueue $t$ on server $j$; $c_s\gets c_s+1$
    \If{$c_s=b_{\Delta}$} \State flush one aggregated delta; $c_s\gets 0$ \EndIf
    \If{\textproc{newCacheAvailable}()} \State refresh local cache \EndIf
    \State \textbf{return} $j$
    \end{algorithmic}
\end{algorithm}

\section{System Design} \label{system_overview}
\begin{figure}[t]
    \centering
    \includegraphics[width=0.75\columnwidth]{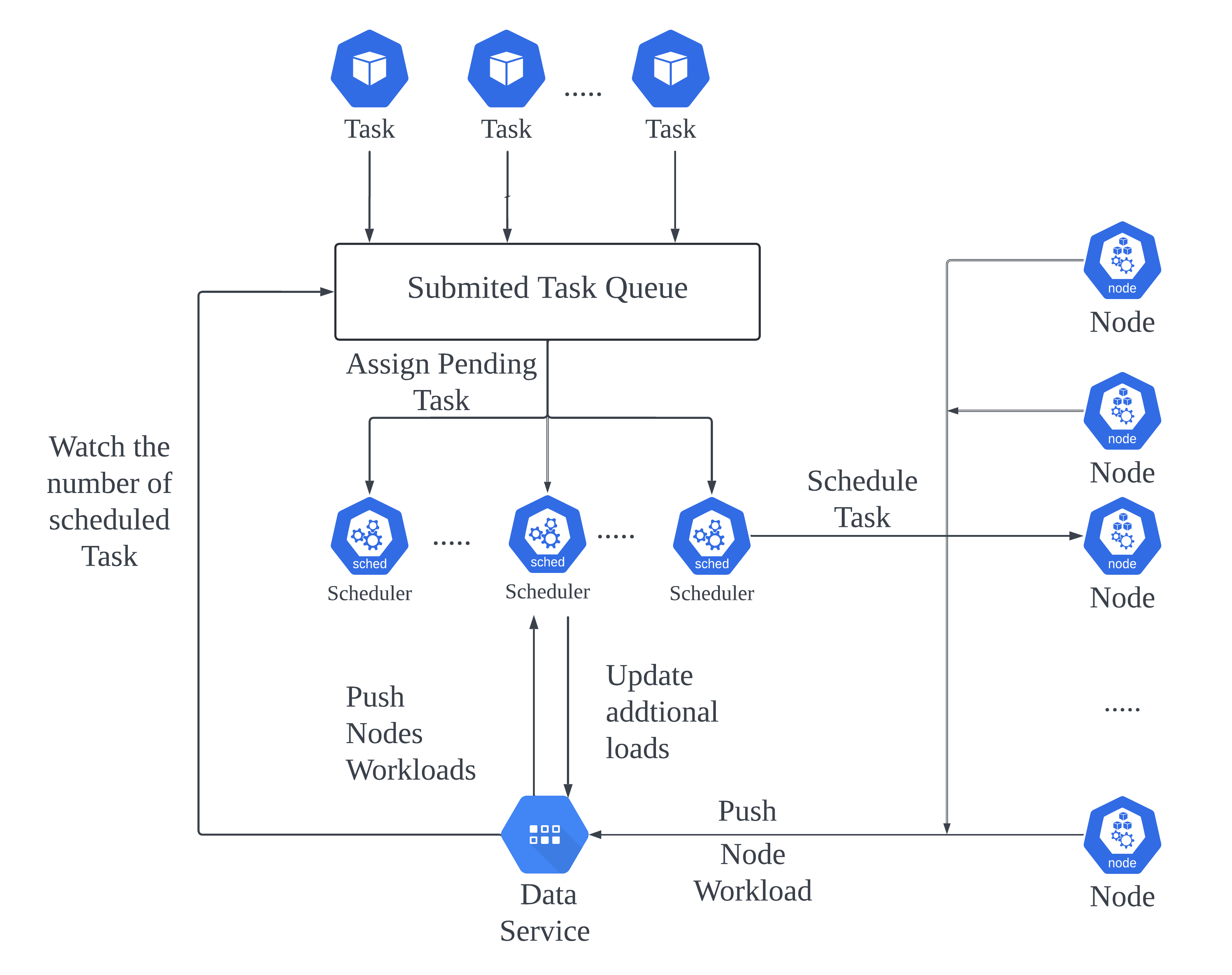}
    \caption{Dodoor architecture}
    \label{fig:dodoor_framework}
\end{figure}

\subsection{Scheduler and Data APIs} \label{section:scheduler_service}
Dodoor’s architecture is shown in Fig.~\ref{fig:dodoor_framework}. Its control path has three roles: client-facing schedulers, worker nodes, and a lightweight data service. The scheduler exposes a submission API and performs scheduling based on cached state, so request handling never blocks on remote probes. Schedulers flush aggregated deltas every local mini-batch of size $b_{\Delta}$, while the data service pushes refreshed cluster snapshots every global batch of size $b$.

This split keeps the hot path short while sharing useful global signals. In practice, each scheduler maintains a local cache keyed by node ID, containing resource-load vectors and queued-duration estimates. Placement reads only this local cache. The $b_{\Delta}$-level aggregation reduces per-request control-plane fanout, while the $b$-level pushes refresh scheduler views without hot-path probing.

\subsection{Server Execution} \label{section:worker}
Each server hosts a worker with a pending-task queue and a concurrency controller. Following standard early-binding, scheduled tasks are admitted only when their requested CPU and memory can be reserved against residual server capacity; otherwise, they remain queued. Tasks are considered in First-Come First-Served (FCFS) order and may execute in parallel up to the node's core-level concurrency limit. If the head-of-queue task cannot be admitted, it blocks until active tasks release sufficient resources.

On completion, the worker reports updated usage and completed duration to the data service. These reports refresh stale cached values and improve subsequent placements. This design preserves decentralized scheduling decisions while letting workers provide low-cost corrective feedback.

\subsection{Cluster Operability, Reliability, and Scalability} \label{section:cluster}
Dodoor tolerates temporary data-service failures by design. Schedulers continue operating on the latest local snapshot and do not require synchronous coordination with peer schedulers. During recovery, the data service rebuilds node state from worker reports and scheduler deltas, then resumes batched pushes.

Scale-out and scale-in operations are handled through node registration and unregistration events. New nodes become ready to be scheduled after the next pushed snapshot. Removed nodes are excluded in the same way. Because updates are batched and idempotent at node granularity, control logic remains simple and robust under churn.

From a scalability perspective, the data service is intentionally off the request-critical path; it acts as an asynchronous aggregation layer rather than a queried database, taking non-blocking writes (scheduler deltas and worker snapshots) and pushing batched state to schedulers. Task dispatch therefore never incurs synchronous reads, eliminating read-lock contention on shared state. Further the batched design explicitly tolerates stale views, so no strict global consistency or scheduler synchronization is required. The current prototype uses a single instance for simplicity. And, the push-based pattern can migrate to an eventually consistent sharded key-value backend without changing scheduler logic.

\section{Implementation} \label{section:implementation}
We implement Dodoor as a modular Apache Thrift prototype in approximately 5,300 lines of Java. The system has four components: a request generator, multiple scheduler replicas, worker agents, and a data service.

Each scheduler uses separate RPC methods for (1) request submission, (2) placement, and (3) batched delta flushing, so bursts in incoming traffic do not block state dissemination. The data service maintains append-friendly in-memory maps behind Thrift APIs, and worker agents track per-task lifecycle transitions in local queues for stable accounting and reproducible measurements.
The implementation exposes a small set of tunable parameters including global push batch size $b$, score weight $\alpha$, and scheduler-side delta flush interval $b_{\Delta}$.

For evaluation, we instrument service-level metrics including end-to-end latency, scheduling latency, RPC message volume, and cluster utilization. To keep randomized comparisons reproducible, we use task ID as the random seed across Dodoor and all baselines. We also implement the decentralized early-binding baselines within the same framework based on their documentation and papers. Dodoor is open-sourced at \url{https://github.com/AKafakA/dodoor}.

\section{Evaluation} \label{section:evaluation}
\subsection{Testbed and Baseline} \label{section:testbed}
Experiments run on 101 CloudLab nodes (100 workers + 1 d6515 control node hosting the data service and 5 independent scheduler processes)~\cite{cloudlab}. The cluster is intentionally heterogeneous across four worker classes, summarized in Table~\ref{tab:cluster-spec}; we focus on CPU and memory dimensions. In our workloads all tasks are admissible on all node types, so heterogeneity affects placement quality through runtime and capacity differences rather than hard feasibility constraints. Each scheduler/data-service process handles up to 8 RPC messages concurrently (up to 40 parallel decisions overall with 5 schedulers in total) and is capped at 16 RPC clients at worker nodes; 100 warm-up requests per experiment are excluded from the reported metrics. The request generator dispatches tasks to the 5 schedulers in round-robin order with Poisson inter-arrivals at the target QPS. Dodoor's default parameters are $b=100$ (cluster size), $\alpha=0.5$ (no bias), and $b_{\Delta}=8$.
\begin{table}[t]
\centering
\caption{Cluster configuration with CPU frequency (101 nodes).}
\label{tab:cluster-spec}
\begin{tabular}{|l|r|r|r|}
\hline
Type & Nodes & CPU (cores @ GHz) & RAM (GB) \\
\hline
d6515 & 1 & 32 @ 2.35 & 128 \\
m510 & 40 & 8 @ 2.0 & 64 \\
xl170 & 25 & 10 @ 2.4 & 64 \\
c6525-25g & 18 & 16 @ 3.0 & 128 \\
c6620 & 17 & 28 @ 2.1 & 128 \\
\hline
\end{tabular}
\end{table}

\begin{figure}[t]
    \centering
    \includegraphics[width=0.85\columnwidth]{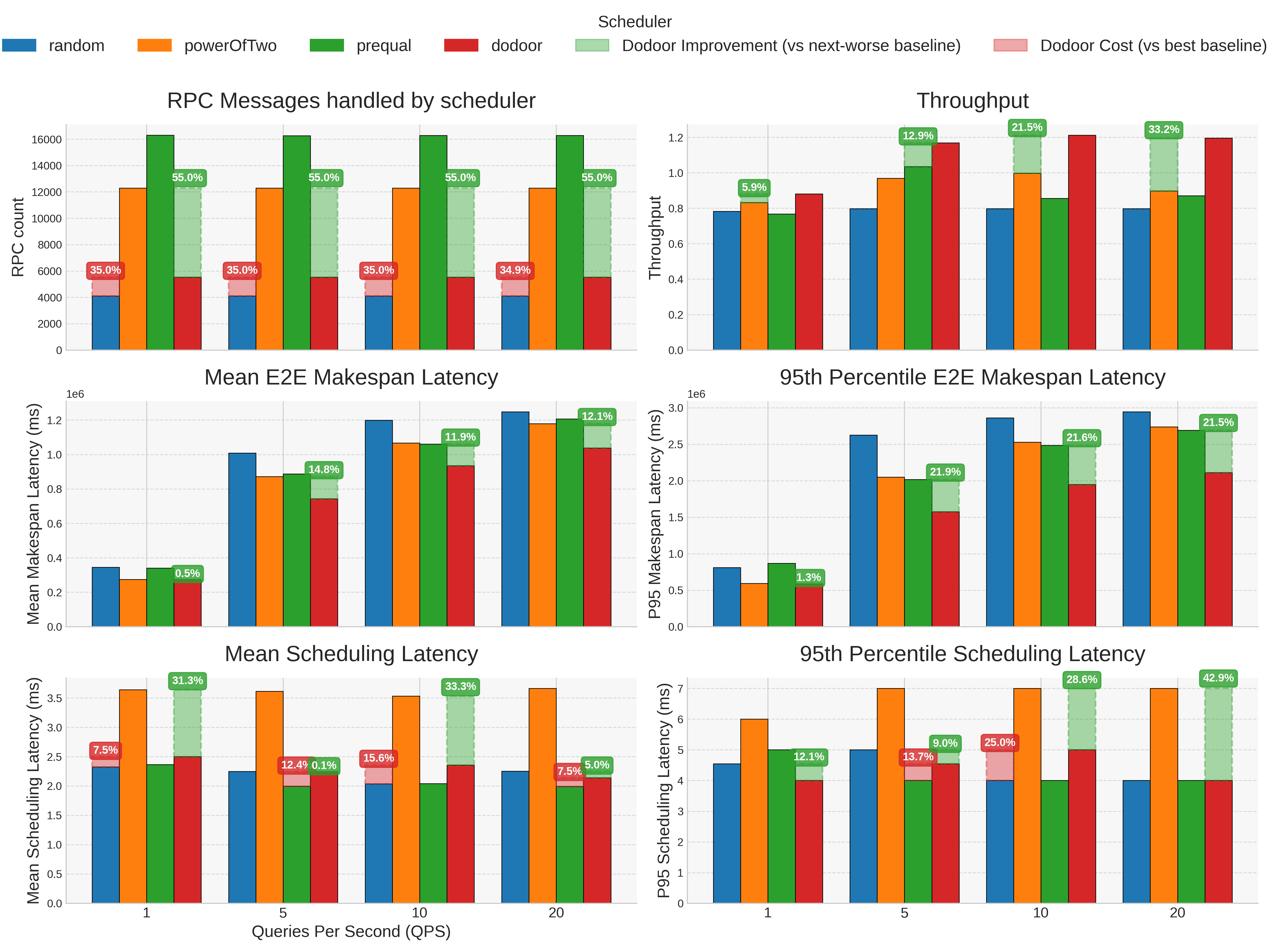}
    \caption{Aggregated scheduling metrics for the Azure workload (x-axis: target QPS).}
    \label{fig:azure_metrics}
\end{figure}

\begin{table}[t]
\centering
\caption{Selected FunctionBench Python tasks}
\label{tab:function-tasks}
\begin{tabular}{|l|r|r|r|}
\hline
Task & CPU cores & Mem (MB) & Runtime (ms) \\
\hline
\texttt{float\_op} (floating-point operations) & 1--2 & 8 & 219--349 \\
\texttt{pyaes} (AES cryptography task) & 1--2 & 9--11 & 222--362 \\
\texttt{linpack} (solves dense system of linear equations) & 4--14 & 29--35 & 372--595 \\
\texttt{matmul} (matrix multiplication) & 4--14 & 37--41 & 456--699 \\
\texttt{chameleon} (3D render) & 2 & 37--38 & 569--966 \\
\texttt{rnn\_name\_gen} (RNN inference) & 4--14 & 467--470 & 1738--3132 \\
\texttt{lr\_predict} (logistic regression inference) & 4--14 & 209--210 & 2462--4341 \\
\texttt{lr\_train} (logistic regression training) & 4--14 & 212--213 & 3532--16201 \\
\hline
\end{tabular}
\end{table}

\begin{figure}[t]
    \centering
    \includegraphics[width=0.85\columnwidth]{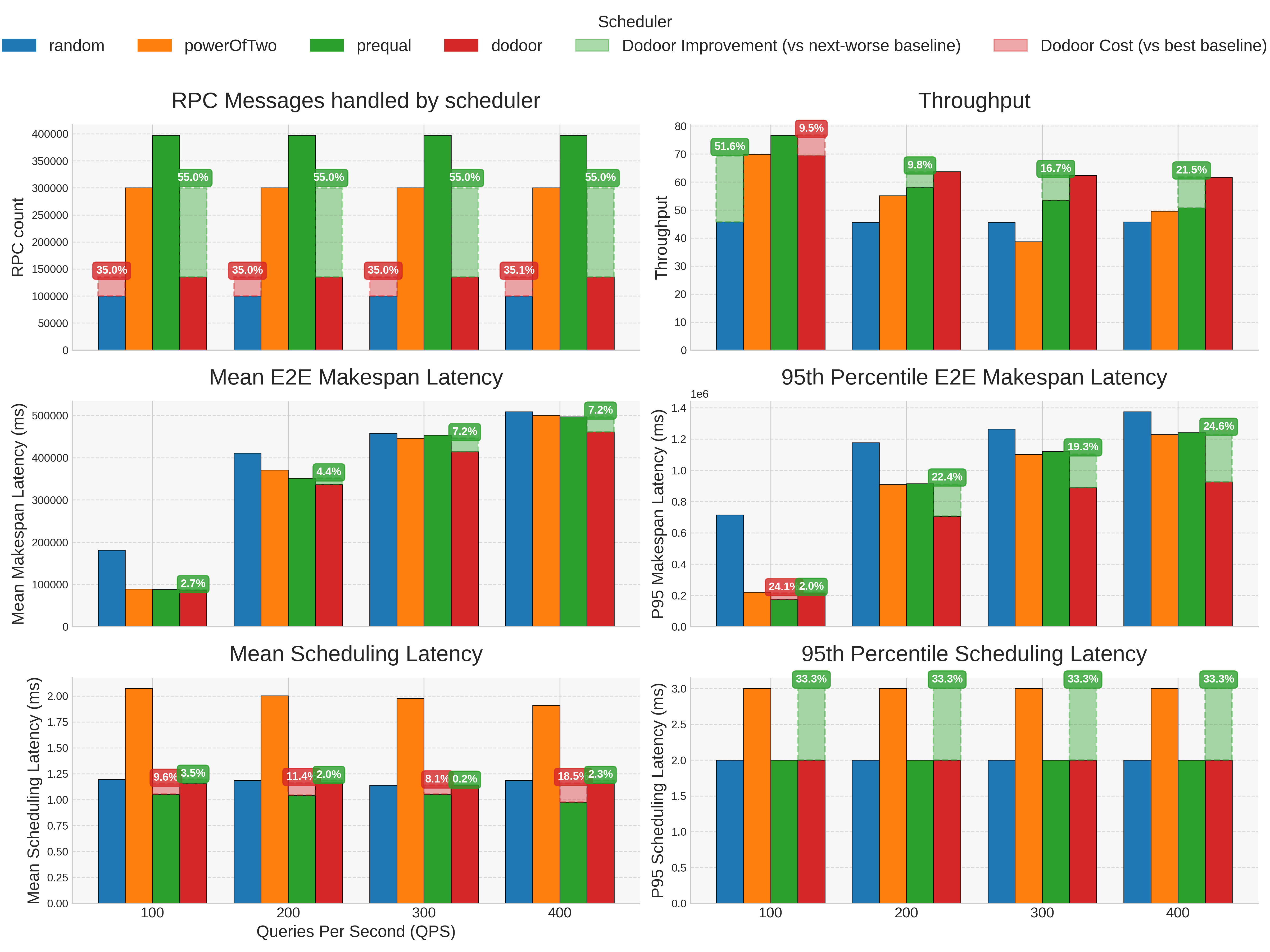}
    \caption{Aggregated scheduling metrics for the FunctionBench workload (x-axis: target QPS).}
    \label{fig:function_metrics}
\end{figure}

We compare Dodoor against three representative decentralized randomized baselines: Random (uniformly selects one feasible server), PoT (standard power-of-two scheduling with online probing before each decision~\cite{envoy,f5NGINXPower}), and Prequal (probe-and-cache with the original recommended parameters $r_{\text{probe}}=3$, $s_{\text{pool}}=16$, $Q_{\text{rif}}=0.84$, $b_{\text{reuse}}=1$, $r_{\text{remove}}=1$~\cite{prequal}).

Our comparison scope focuses on decentralized early-binding schedulers. Fully centralized schedulers that assume continuously consistent global state represent a different design point and are outside our comparison scope. We do not include least-loaded cached variants since deterministic best-candidate selection under stale replicated snapshots can amplify herding on apparently light nodes, whereas PoT preserves controlled randomization under decentralized parallel operation.

Following Prequal's discussion that Sparrow's messaging overhead is high for this class of systems, we omit Sparrow from our quantitative baselines. More fundamentally, as detailed in Section~\ref{section:variants_of_random}, Sparrow's late-binding protocol is primarily designed and evaluated in homogeneous slot-based settings. Under multidimensional constraints, the reservation-to-commit gap can increase confirmation races under concurrent stale views. We therefore focus on PoT and Prequal, which are more directly comparable to Dodoor's decentralized early-binding setting, and leave exploration of heterogeneous late binding for future work.

\subsection{Trace-Replay Evaluation Under VM Placement} \label{section:azure}
We replay 4,000 Azure VM requests with heterogeneous CPU/memory demands and durations~\cite{AzureCloud}; requests are filtered to fit cluster capacities and to have durations below 10 minutes. Each request is mapped to a \texttt{stress-ng} job with matched CPU/memory pressure and target duration. The target duration is used as the scheduler estimate $\hat{d}$, while actual wall-clock \texttt{stress-ng} durations still deviate from this estimate due to system noise and contention. Compared with PoT and Prequal baselines, Dodoor consistently reduces control-plane communication and improves placement quality at medium-to-high load (Fig.~\ref{fig:azure_metrics}).

Across QPS settings, Dodoor reduces scheduler RPC messages by about 55\% versus PoT and 66\% versus Prequal. Throughput is computed as completed requests divided by experiment wall-clock time, and scheduling latency is end-to-end makespan minus server-side tracked execution lifetime. Dodoor improves throughput by 5.9\%--33.2\% and lowers mean and P95 end-to-end latency by up to 14.8\% and 21.9\%, respectively. Even under low load we observe consistent latency gains (0.5\% mean, 1.3\% tail), so the cached resource-aware score helps both lightly and heavily loaded regimes (Fig.~\ref{fig:azure_metrics}).

\subsection{Container-based Evaluation for Computing Tasks} \label{section:function-bench}
We run short-lived heterogeneous Python functions in Docker as FunctionBench-style tasks~\cite{functionBench}. Before online scheduling, we perform offline profiling per (task type, server class) to obtain the a-priori duration estimate $\hat{d}$; during online runs, actual durations deviate from this estimate under concurrent load, reflecting non-oracle conditions. The replay set contains 8 task types covering numerical kernels, inference/training, and utility workloads with diverse runtime/resource profiles. Table~\ref{tab:function-tasks} reports task demands and runtime ranges across node types; aggregated online results are in Fig.~\ref{fig:function_metrics}.

At low load (QPS=100), Dodoor already improves mean/P95 end-to-end latency by 2.7\%/2.0\% over PoT. Under cluster saturation, throughput rises by up to 21.5\%, mean latency drops 7.2\%, and tail latency drops 24.6\% (Fig.~\ref{fig:function_metrics}); communication gains remain similar to Azure. Across both workloads, throughput and latency improve jointly rather than trading off, because removing hot-path probing returns control-plane capacity to task execution while the resource-load score curbs queue buildup that would otherwise inflate the tail; the residual freshness-vs.-overhead lever is the batched-refresh size $b$ (Section~\ref{section:sensitivity}).

\subsection{Sensitivity Analysis} \label{section:sensitivity}
\begin{figure}[t]
    \centering
    \includegraphics[width=0.85\columnwidth]{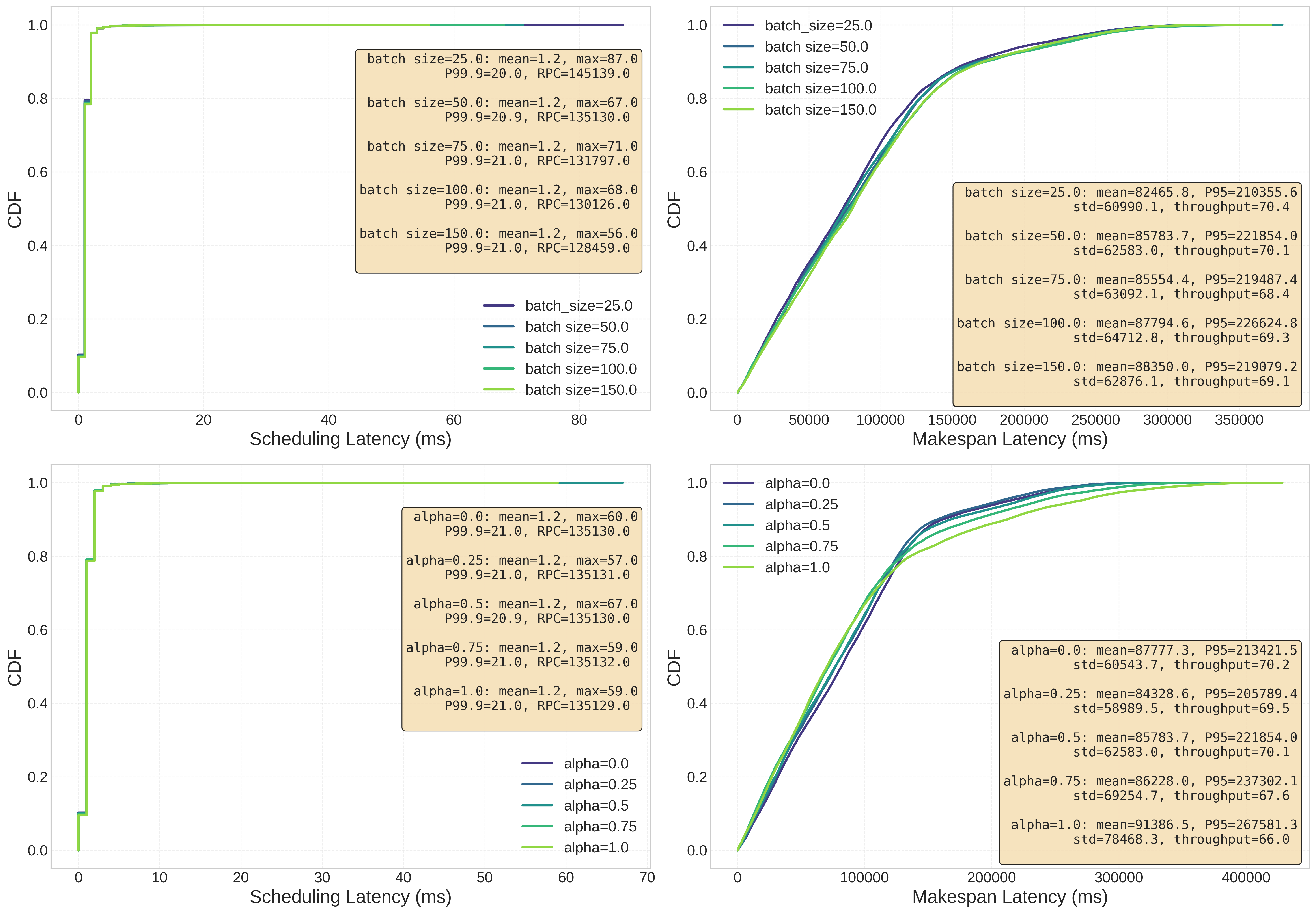}
    \caption{Sensitivity to batch size $b$ and score weight $\alpha$ (FunctionBench, QPS=100).}
    \label{fig:parameter_tuning}
\end{figure}

We vary batch size $b\in\{25,50,75,100,150\}$ and observe the expected freshness/overhead tradeoff (Figure~\ref{fig:parameter_tuning}). Smaller batches provide fresher cache views and improve placement quality, reducing mean makespan by up to 7.1\%. However, frequent pushes increase scheduler load: in our setting, handled messages rise to 16,680 and maximum scheduling latency reaches 31~ms when $b$ is smallest. This batch-size sensitivity also provides a first-order view of Dodoor's behavior under rapidly changing workloads: smaller batches improve freshness at the cost of higher control-plane overhead.

We also vary score weight $\alpha\in[0,1]$, which controls resource-balance versus queue-duration emphasis in Algorithm~\ref{alg:scheduling}. Lower $\alpha$ yields tighter tail distributions by prioritizing multidimensional resource fit. As $\alpha$ approaches 1, latency variance increases and tail behavior degrades, with $\alpha=1$ giving the worst case because resource contention is effectively ignored.

\section{Limitations and Applicability}
Dodoor targets decentralized early-binding scheduling in heterogeneous clusters with parallel schedulers and cached state. Our evaluation covers CPU and memory dimensions and excludes accelerators, network-aware placement, and fault-injection experiments. Rather than globally optimal placement, we target low overhead and robust performance under periodically refreshed stale views; Section~\ref{section:sensitivity} quantifies the freshness-overhead tradeoff. Dodoor's modular scoring and push-based state interface make extension to additional resource types and fault scenarios a straightforward next step.

\section{Related Work}
Prior systems span centralized global-view schedulers and decentralized randomized schedulers. Centralized designs improve coordination through richer shared state~\cite{quincy,borg,Tetrisched,omega,apollo}, while decentralized designs improve scalability via lightweight sampled decisions~\cite{envoy,f5NGINXPower,hawk,eagle,prequal,Murmuration}. Probe-reduction methods reduce per-request communication but expose a freshness-overhead tradeoff under parallel schedulers~\cite{sparrow,prequal}. Stale/batched analyses and multi-resource placement studies further show that robust decisions must handle delayed state refresh and multidimensional fit~\cite{stale_load,batchSetting,power_of_two_batches,tetris,head-of-line-tcp}. Dodoor follows this direction with batched cache refresh and heterogeneity-aware decentralized scoring.

\section{Conclusion}
\label{sec:conclusion}
Dodoor is a decentralized scheduler for heterogeneous clusters that reduces probing overhead while improving placement quality. By combining batched cache updates with a resource-load score, Dodoor achieves substantial communication reduction and strong latency and throughput performance on real-cluster runs. Results show that on the decentralized hot path, removing per-request probing can matter more for end-to-end performance than focusing on the last increment of placement optimality. Further, our lightweight multidimensional resource-load score with bounded-staleness batched refreshes improves tail latency without central coordination, while the $b$ knob makes the freshness-versus-overhead tradeoff explicit and tunable.
\begin{credits}
\subsubsection{\ackname}
The authors thank the CloudLab team for providing the experimental testbed used in this work.

\subsubsection{\discintname}
The authors have no competing interests to declare that are relevant to the content of this article.
\end{credits}

% ---- Bibliography ----
%
% BibTeX users should specify bibliography style 'splncs04'.
% References will then be sorted and formatted in the correct style.
%
\bibliographystyle{splncs04}
\bibliography{bibe}
\end{document}